\documentclass[journal]{IEEEtran}
\ifCLASSINFOpdf
   \usepackage[pdftex]{graphicx}
\else
\fi
\usepackage{amsmath}
\usepackage{amsfonts}
\usepackage{algorithmic}

\usepackage{array}
\usepackage{fixltx2e}
 \usepackage{dblfloatfix}

\usepackage{enumitem}

\usepackage{booktabs} 
\newcommand{\ra}[1]{\renewcommand{\arraystretch}{#1}}

\begin{document}
%
\title{Classification of  syncope through data analytics}
%
%
%

\author{Joseph Hart,
        Jesper Mehlsen,
        Christian H. Olsen,
        Mette Sofie Olufsen,
        and~Pierre~Gremaud
\thanks{P. Gremaud, J. Hart and M. Olufsen are with the Department of Mathematics, North Carolina State University, Raleigh, NC 27695, USA  (e-mail: gremaud@ncsu.edu).}
\thanks{J. Mehlsen and C. Olsen are with the Frederiksberg Hospital, Frederiksberg, Denmark.}
\thanks{This work was supported in part by  the National Institutes of Health and the National Science Foundation under Grant  DMS 1557761 and the National Institutes of Health through grant NIH 5P50GM094503-06 VPR sub-award to North Carolina State University.}}

%
%

\markboth{Journal of \LaTeX\ Class Files,~Vol.~14, No.~8, August~2015}%
{Shell \MakeLowercase{\textit{et al.}}: Bare Demo of IEEEtran.cls for IEEE Journals}
%



\maketitle

\begin{abstract}
Objective: Syncope is a sudden loss of  consciousness with loss of postural tone and spontaneous recovery; it is a common condition, albeit one that is challenging to accurately diagnose. Uncertainties about the  triggering
mechanisms and their underlying pathophysiology have led to various classifications of patients exhibiting this symptom.
This study presents a new way to classify  syncope types using machine learning. {\em Method:} we hypothesize that syncope types can be characterized by analyzing blood pressure and heart rate time series data obtained from the head-up tilt test procedure. By  optimizing classification rates, we identify a small number of  determining markers which enable data clustering. {\em Results:} We apply the proposed method to  clinical data  from 157 subjects; each subject was identified by an expert as being either healthy or suffering from  one of three conditions: cardioinhibitory syncope,  vasodepressor syncope and postural orthostatic tachycardia. Clustering confirms the three disease groups and identifies two distinct subgroups within the healthy controls. {\em Conclusion:} The proposed method provides evidence to question current syncope classifications;  it also offers means to refine them. {\em Significance:} Current syncope classifications are not based on pathophysiology and  have not led to significant improvements in patient care. It is expected that a more faithful classification will facilitate our understanding of the autonomic system for healthy subjects, which is essential in analyzing pathophysiology of the disease groups.

\end{abstract}


%
\IEEEpeerreviewmaketitle

\section{Introduction}
%
%
%
%
\IEEEPARstart{S}{yncope} is defined as a ``transient loss of consciousness due to transient global cerebral hypoperfusion characterized by rapid onset, short duration and spontaneous complete recovery" \cite{defsync}; it is a prevalent disorder which accounts for over 1 million visits to emergency departments per year in the US alone  \cite{probst15}.
 Cerebral hypoperfusion is usually caused by a decrease in systolic blood pressure which, in turn, is linked to a reduction in cardiac output and total vascular resistance; a fall in either can cause syncope, but a combination of both mechanisms is often present \cite{guy,robertson}. 
 Standard diagnostic methods such as  the head-up tilt (HUT) test, discussed below,  only provide information about the integrated cardiovascular response  via measurements of arterial blood pressure (BP) and heart rate (HR). 

Common types of  syncope have been classified to facilitate diagnosis and treatment \cite{brignole,mkm,guidelines,dijk}. 
However, the current classifications are phenomenological and the corresponding terminology is  inconsistent \cite{brignole,dijk}; therapeutic approaches based on them have generally not lead to notable improvements in patients' condition. We concentrate on three patient groups, namely cardioinhibitory syncope, vasodepressor syncope  and postural tachycardia, which are discussed in the next section.

Patients are examined after repeated episodes of lightheadedness and fainting. Even among patients diagnosed with  syncope, these conditions cover a wide range of diseases and are  difficult to diagnose \cite{dijk}. Diagnosis is typically based on patient symptoms along with visual analysis of simultaneous measurements of BP and HR recorded during a postural challenge, most commonly, HUT. For the considered patients, the end result is a significant drop in BP with or without changes in HR; what distinguishes the groups is how these signals change in response to the postural challenge.

In this paper, we  analyze data from subjects referred to a large regional medical center in Copenhagen, Denmark. These subjects present  symptoms of dizziness and fainting--primarily in the upright position--and thus are suspected of syncope associated with autonomic dysfunction.  Our central hypotheses are that (1) syncope etiology can be determined by analysis of BP and HR data and that (2) machine learning and mathematical modeling can fundamentally improve diagnosis accuracy for patients suffering from syncope associated with autonomic dysfunction.

\section{Data and Methods}
\subsection{Head-up tilt test}
This study analyzes data from 157 subjects who have been exposed to a head-up tilt test to examine their ability to control BP and HR. Data were collected between  2004 to 2015 and involve patients admitted to Frederiksberg Hospital, Denmark, after experiencing episodes of syncope as well as a group of healthy control subjects.  Analyzed data  are from subjects  with no known heart or vascular diseases. All data are extracted from existing patient/control records and assigned random identifiers before  analysis.

 \begin{figure*}[h]
 \begin{center}
\includegraphics[width=.34\textwidth]{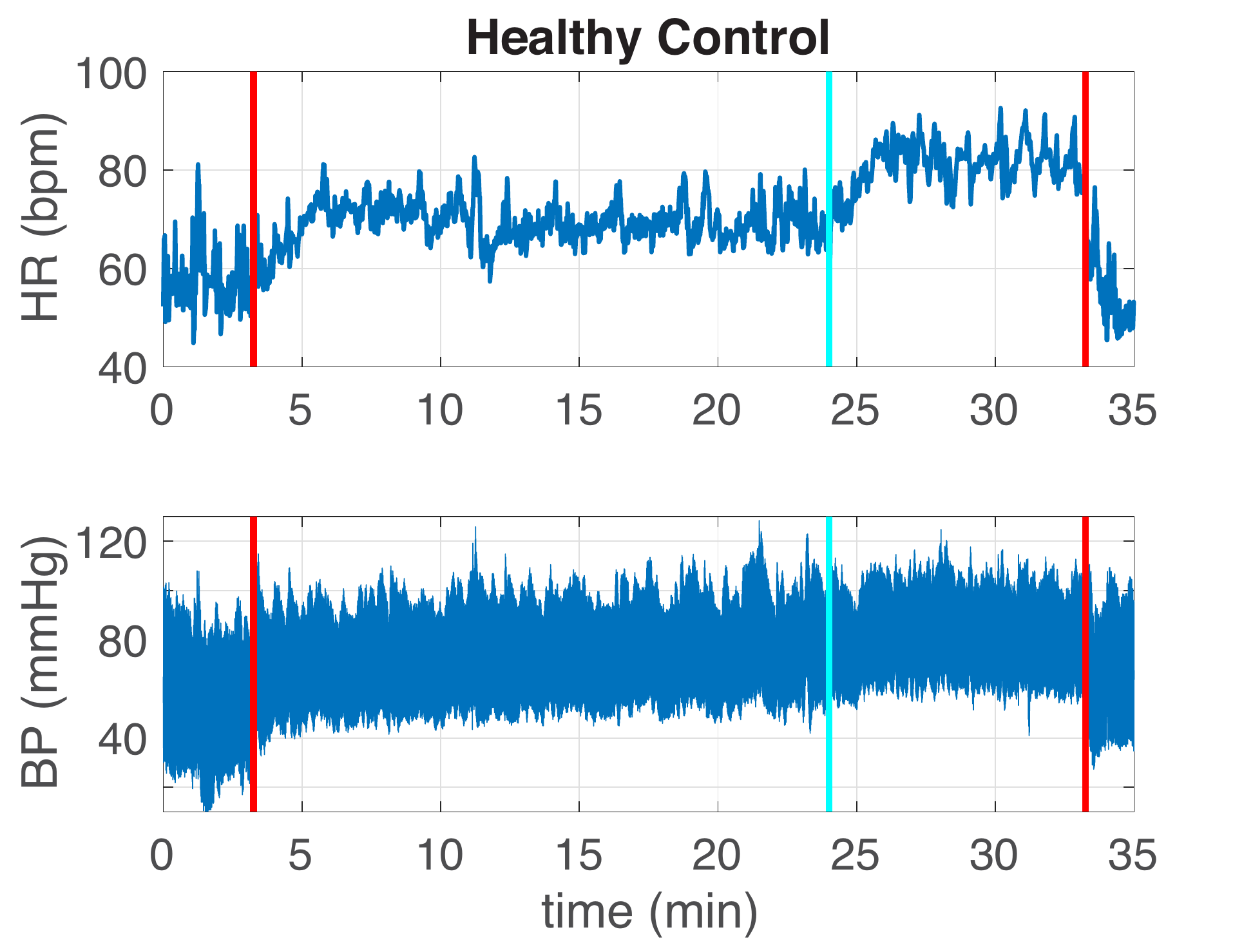} 
\includegraphics[width=.34\textwidth]{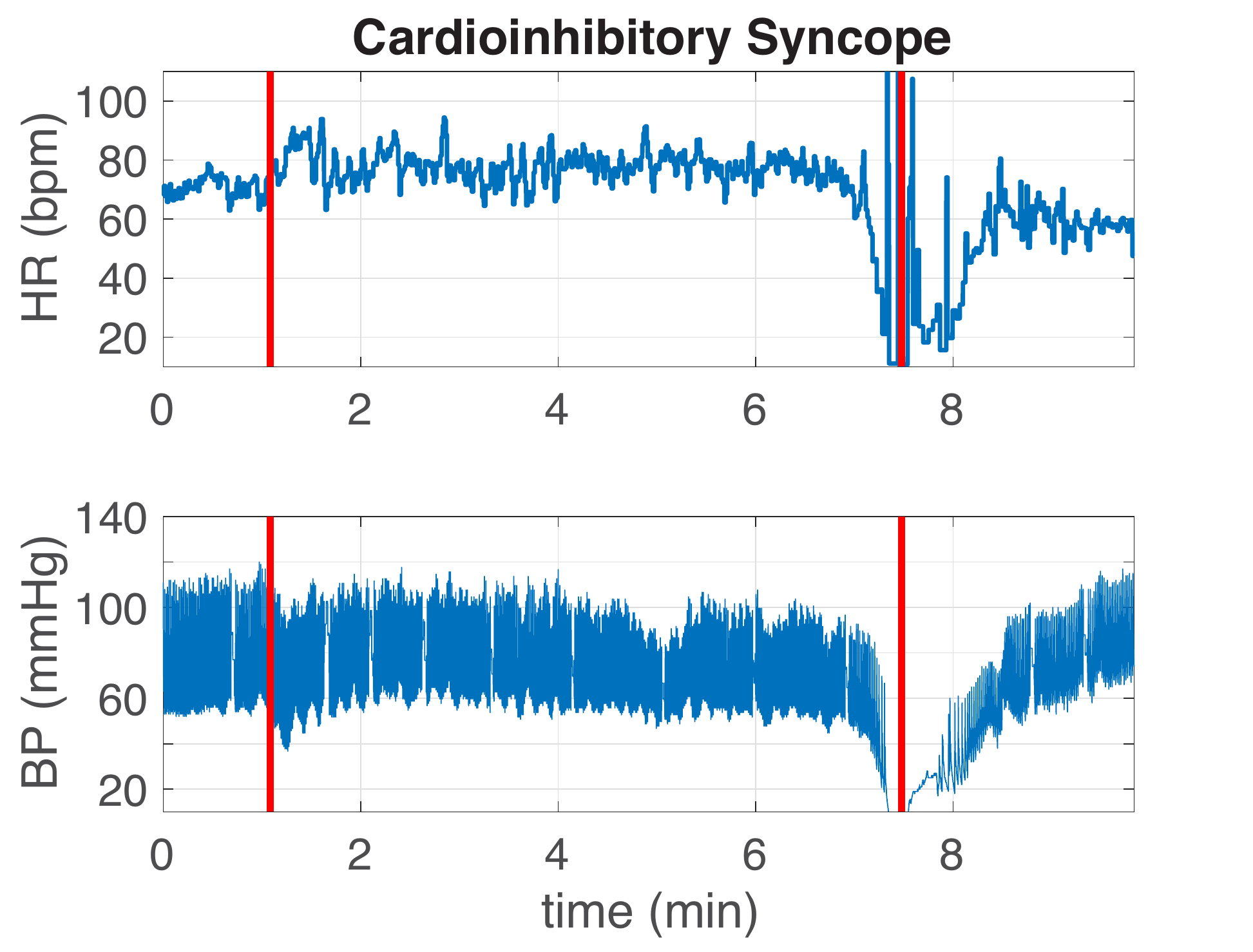} \\
 \includegraphics[width=.34\textwidth]{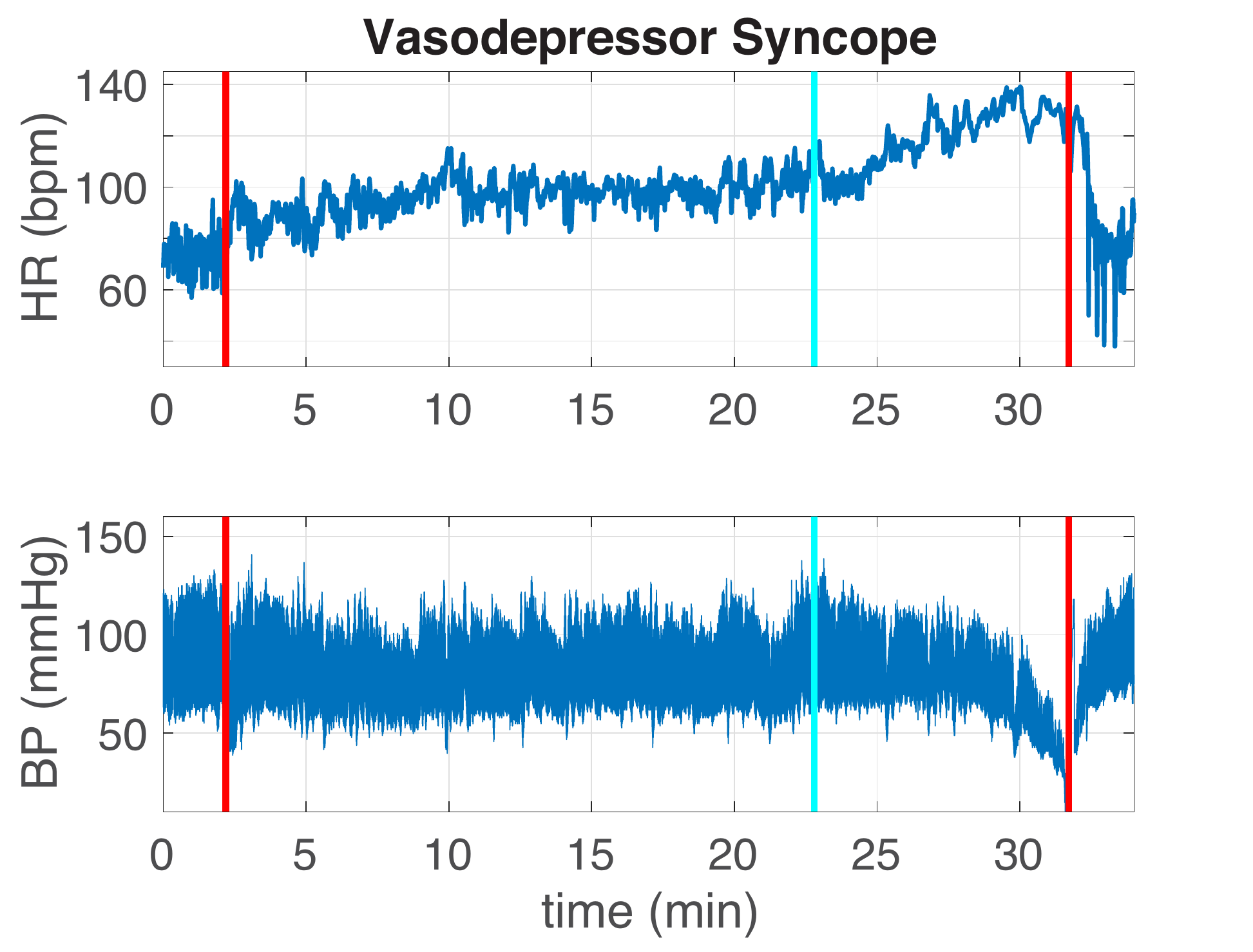} 
\includegraphics[width=.34\textwidth]{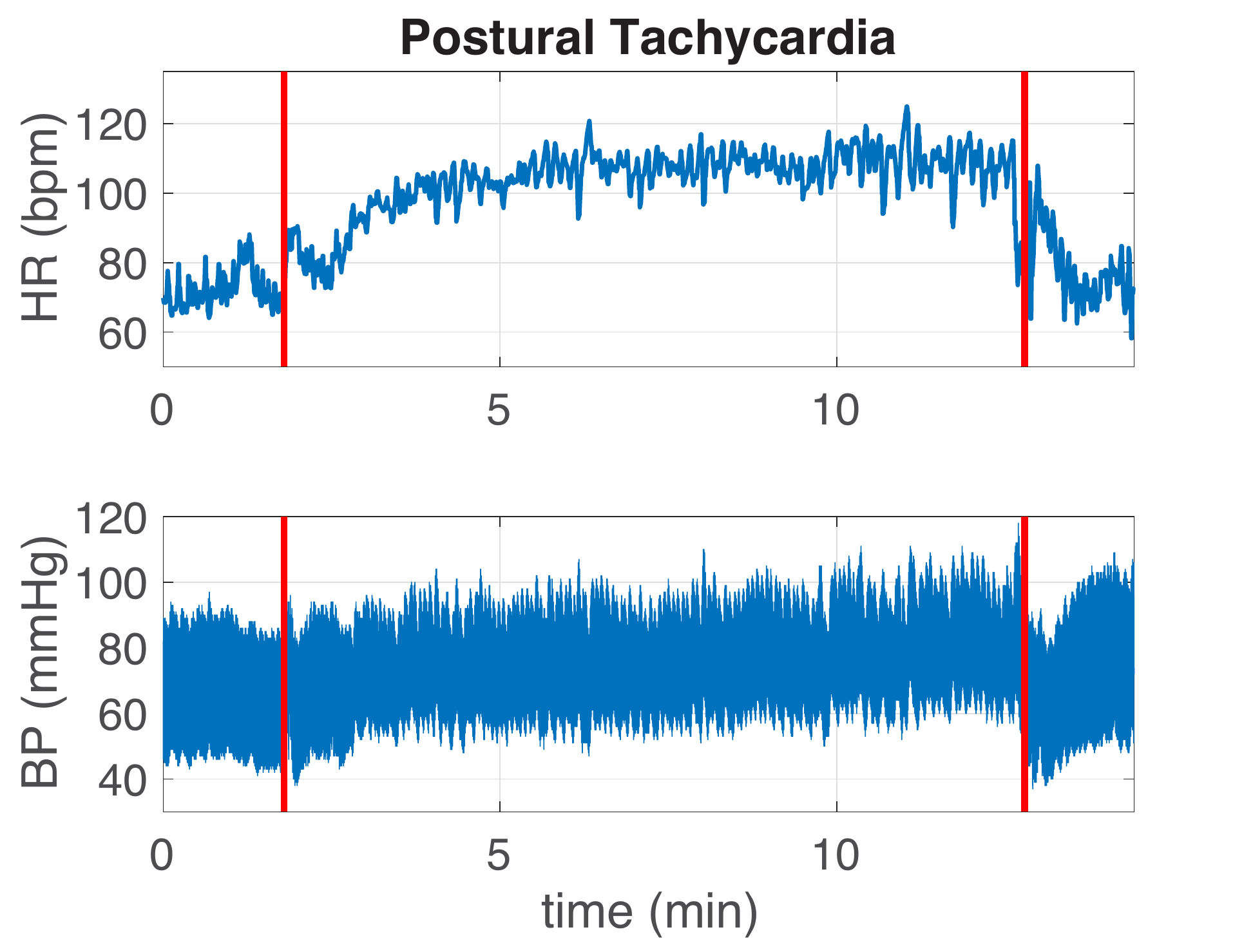} 
\caption{Typical data from the head-up title test for four subjects: healthy control and patients suffering from cardioinhibitory syncope, vasodepressor syncope and postural tachycardia. The redline lines denote the start and end of the tilt from supine position, up to about 60$^o$ and back to supine position. Heart rate (beats per minute) and blood pressure (mmHg) are displayed as functions of time. The blue lines correspond to the administration of nitroglycerine, a vasodilator.}
\label{fig:cases}
\end{center}
\end{figure*}

After arriving at the hospital, all subjects are instrumented with BP and ECG sensors. BP is measured using photoplethysmography (Finapres Medical Systems B.V.) in the index finger of the non-dominant hand. The hand is placed in a sling at the level of the heart. ECG is recorded using standard precordial leads. Continuous ECG and BP signals are sampled a rate of 1.0 kHz and saved digitally using an A/D-converter communicating with a computer via LabChart 7 (ADInstruments). This program allows extraction of HR from the ECG measurement.  After clear signals are detected, the patients rest for 10 minutes in the supine position before being tilted head-up to an angle of 60 degrees at a speed of 15 degree/second measured by way of an electronic marker. The subjects remain tilted head-up during initial passive phase of the test. In case of a negative passive phase, a provocative drug--nitroglycerine--is administered to facilitate the occurrence of a vasovagal reflex. This step is taken after around 20 minutes for the healthy controls and after a variable amount of time for  
 the patients from the other groups, see Fig.~\ref{fig:cases}.  Patients are returned to the supine position at the same tilt speed after a total of 30 minutes or earlier if they present  signs of syncope or presyncope.

 \subsection{Data and clinical classification}

 For each subject,   time series measurements of HR and BP are available  over the course of the head up tilt test. Our analysis is based on data starting at  two minutes before the tilt up and lasting until two minutes after the tilt down. The duration of the test varies for each subject and thus so do the lengths of the time series. We denote by  $p_i$ the number of samples taken for subject $i$, $i=1, \dots, 157$. The time series data for the $i$-th subject  have the form 
\begin{eqnarray*}
 h^i&=&(h_1^i,h_2^i,\dots,h_{p_i}^i),  \\
 b^i&=&(b_1^i,b_2^i,\dots,b_{p_i}^i),
 \end{eqnarray*}
 where $h$ and $b$ stand  respectively for HR and BP.  Each subject has been identified by a clinician as either {\em healthy} or suffering from  {\em cardioinhibitory syncope}, {\em vasodepressor syncope} or {\em postural orthostatic tachycardia (POTS)}, see Fig.~\ref{fig:cases} and text below. The corresponding distribution of subjects is given in Table~\ref{tabdata}. When administered, nitroglycerine was given sublingually at a dose of 0.4 mg;  it was given to  94\% of the healthy controls and to, respectively, 64\%, 78\% and 0\%  of the cardioinhibitory syncope, vasodepressor syncope and POTS patients.

\begin{table}[h]
\centering
\ra{1.3}
\begin{tabular}{lcccc}
\toprule
class  & subjects & age range & age mean/median & \% female\\
\midrule
healthy & 89 & 14--92 & 50/49 & 67 \\
cardio. syn.  & 28 & 15--80 &  33/31 & 63 \\
vasodep. syn. & 27 & 67--91 & 58/63 & 67 \\
POTS  & 13 & 16--38 & 24/22 & 85 \\
\bottomrule
\end{tabular}
\vskip.2cm
\caption{Summary of subject distribution.} 
\label{tabdata}
\end{table}

{\em Cardioinhibitory syncope}  results from excessive pooling of blood in the lower extremities. In response, the Bezold Jarish reflex stimulates the vagal nerve decreasing HR, and subsequently BP, leading to syncope. Subjects in this group experience 
only a few and if any pre-syncope symptoms. {\em Vasodepressor syncope} also leads to fainting due to excessive pooling of blood in the extremities. This condition has a longer time scale for the fall in BP allowing prominent pre-syncopal symptoms. For these patients, the Bezold Jarish reflex likely inhibits sympathetic vasoconstriction thus resulting in a significant drop in BP, which may or may not be followed by a drop in HR, eventually inducing syncope. Finally, patients experiencing POTS may have a reduced central blood volume causing BP regulation to be challenged by changes in intrathoracic pressure due to respiration. This causes pathological fluctuations in BP with phase-shifted changes in HR elicited by the baroreceptor control system. In particular, the patients in this group have excessive vagal withdrawal leading to inappropriate increases in HR, which further reduces cardiac filling due to a shortening of the diastolic filling time. As a result, HR increases while BP oscillates \cite{pots}. In addition to these three patient groups, we analyze data from a large group of healthy controls. These  subjects were admitted as described above but had a normal outcome during testing of the autonomic nervous system.  
Diagnosis of patients was done by one of the coauthors (Mehlsen) based on data from the test analyzed here, on spontaneous HR variability, and on  knowledge of general symptoms and signs displayed (not used in this study).

 The  classification corresponding to the above expert diagnosis is denoted $Y^i$, $i=1,\dots, 157$, where, for any subject, $Y$ takes values in the four classes introduced in the previous paragraph. The complete data is thus 
 \[
 \{ h^i, b^i, Y^i\}_{i=1}^{157}.
 \]

\begin{figure}[t]
\centering
\includegraphics[width=2.9in]{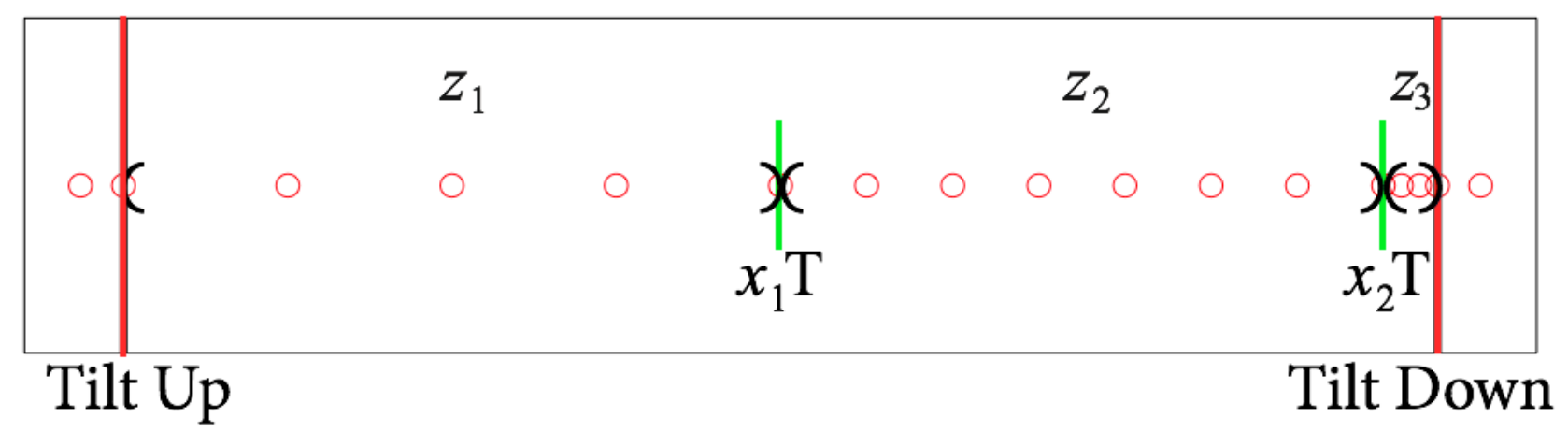}
\caption{Marker parameterization: the red lines represent tilt up and tilt down times, separated by an elapsed time of $T$; that interval is split into three subintervals $[0, x_1T], [x_1 T , x_2 T]$ and  $[x_2 , T]$, see green lines. Each subinterval contains $z_i$ nodes, $i=1,2,3$. The nodes (sample times) are illustrated by the red circles. Optimal numerical values of these parameters are given Table~\ref{tabopti}.}
\label{fig:parameterization}
\end{figure}

%
The time series are first subsampled at 20Hz, down from 1000Hz in the original signal. Second,  the signals are preprocessed through  a moving average window with a width of 1000 points (or equivalently 50 seconds). Finally,  we normalize each signal by subtracting its global mean for each subject. We denote the preprocessed normalized time series by ${\cal H}^i$ and ${\cal B}^i$ where 
\[
({\cal H}^i, {\cal B}^i) \in {\mathbb R}^{N^i}\times {\mathbb R}^{N^i}, \qquad  i=1,\dots,157,
\]
with $N^i$ referring to the number of retained sample values for the $i$-th subject.

\subsection{Random Forest classifier}

A Random Forest \cite{breiman2001} \cite{hastie2009} is an ensemble of classification trees \cite{breimanbook}; this method has proven to be successful in a variety of fields  \cite{diaz, cutler, pal}. We use its implementation in the R {\sc{randomForest}} function. To avoid overfitting and improve model performance, the models are learned not on the full dataset $({\cal H}^i, {\cal B}^i)_{i=1}^{157}$ but on a lower dimensional set of features (or makers) extracted from the data. 
We show below that high classification rates can be obtained by restricting these markers to simple time sampling of both the normalized HR and BP signals $({\cal H}^i, {\cal B}^i)_{i=1}^{157}$. 

For each subject, one marker is placed one minute before the tilt up and an other one minute after the tilt down. 
We parameterize the placement of the  remaining  markers by partitioning the ``tilt up to tilt down  interval''  into three subintervals
\[
 [0, x_1T_i], [x_1 T_i , x_2 T_i] \mbox{ and }[x_2 T_i, T_i],
 \]
  where $0<x_1<x_2< 1$ and $T_i$ denotes the elapsed time  between tilts for the $i$-th subject. Further, we consider as potential markers  $z_j$ points uniformly spaced in the $j$-th subinterval, $j=1,2,3$. For each interval, we retain the sampled values which are the closest in time to
 \begin{description}[align=right,labelwidth=3cm]
 \item [interval 1:]  $T^{up} + \ell\, \frac {x_1}{z_1-1}T$, $\ell=0,\dots, z_1-1,$
 \item [interval 2:]  $T^{up}+  (x_1  + \ell\, \frac {x_2-x_1}{z_2})T$, $\ell=1,\dots, z_2,$
 \item [interval 3:]  $T^{up} + (x_2  + \ell\, \frac {1 -x_2 }{z_3})T$, $\ell=1,\dots, z_3,$
 \end{description}
  with the additional conventions that if $z_1 = 0$, there is no node in the first interval and if $z_1 = 1$, the first interval only contains the node corresponding to the tilt up time, $T^{up}$.
  This parameterization is illustrated  in Fig.~\ref{fig:parameterization}.

\begin{figure}[h]
\centering
\includegraphics[width=2.5in]{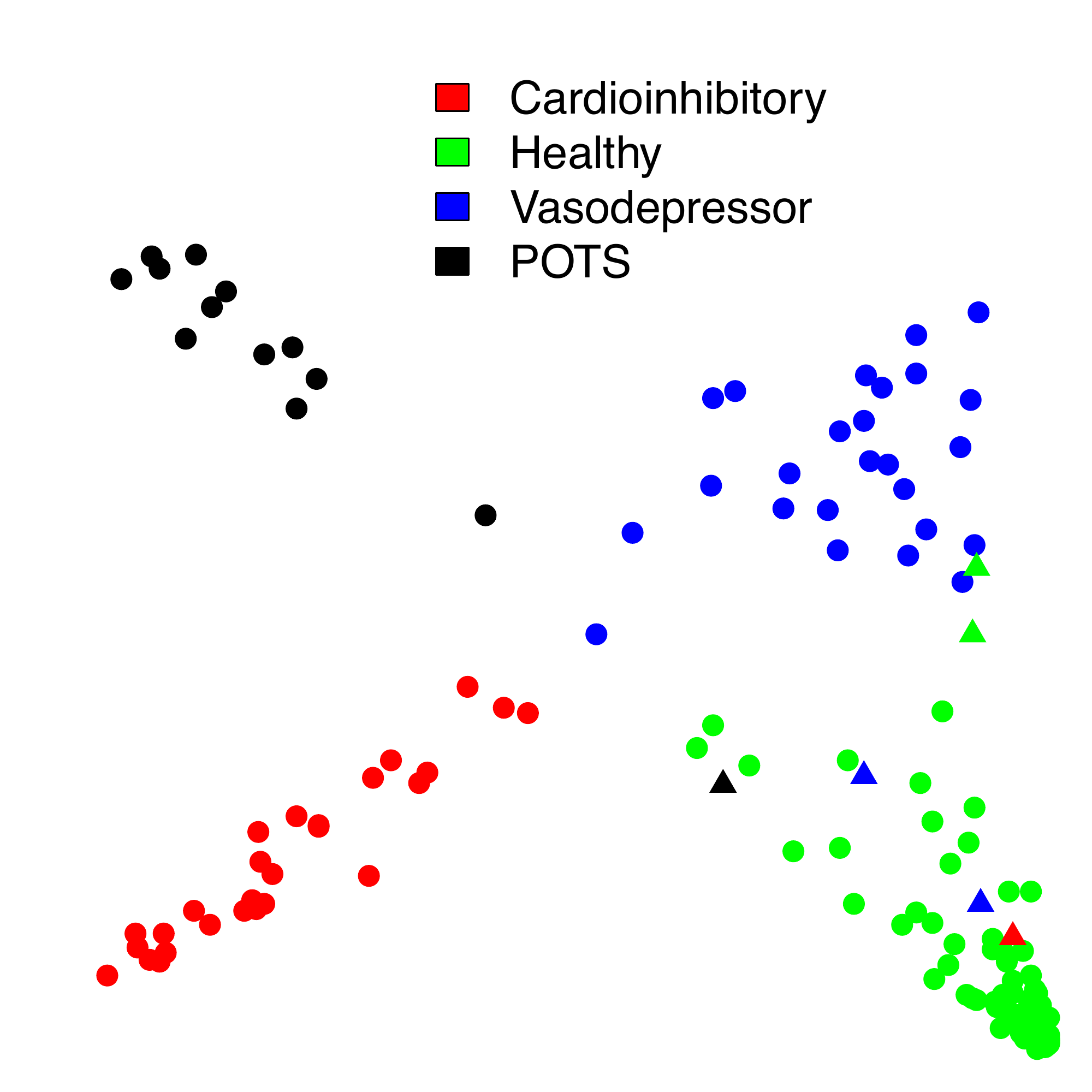}
 \caption{Barycentric coordinate representation of the classification of four classes. Misclassified subjects are denoted by a $\bigtriangleup$. The classification  is 96\% successful. Point tightness  indicates how well-defined a specific class is.}
\label{figclass}
\end{figure}

 \begin{figure*}[t]
 \begin{center}
\includegraphics[width=2.3in]{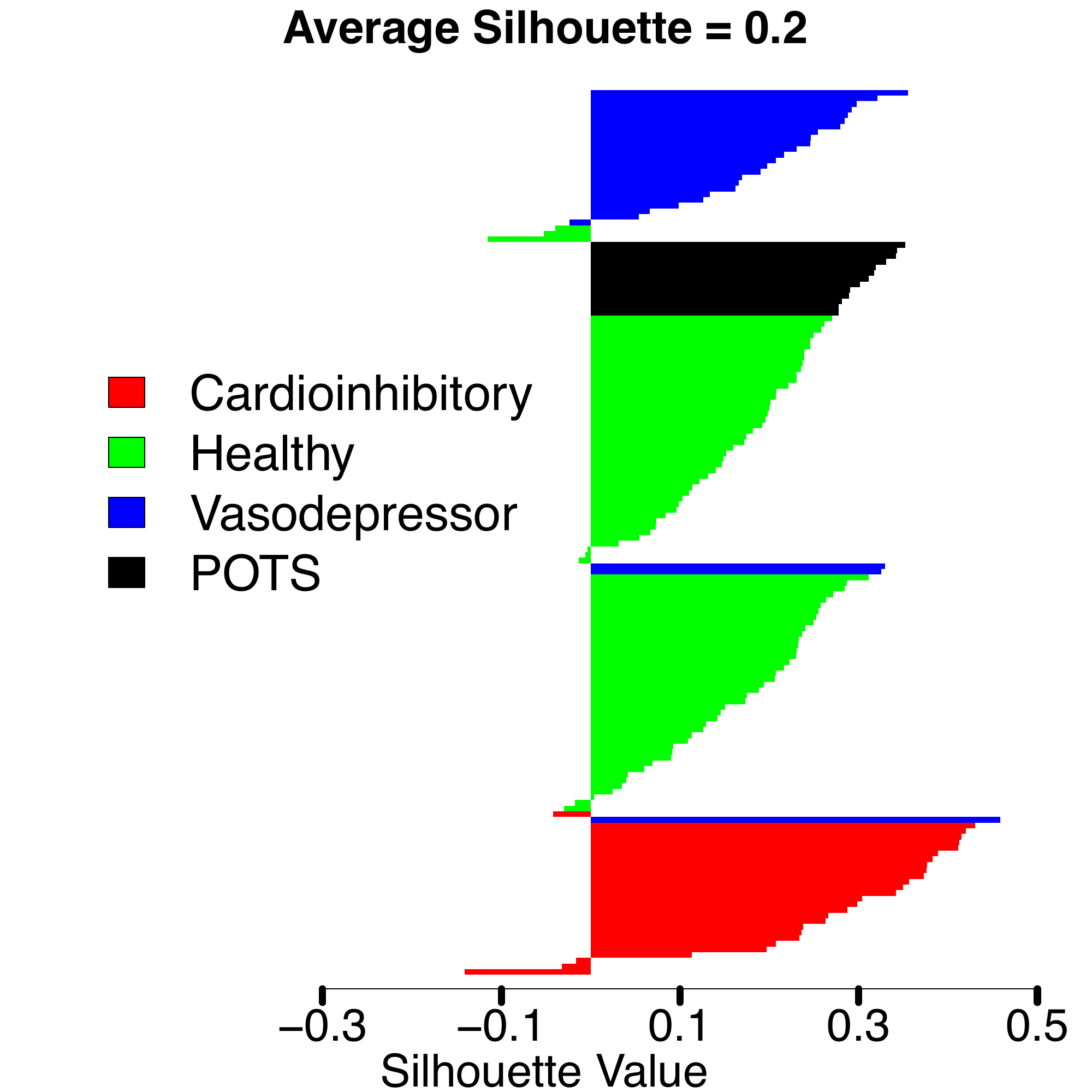}
\includegraphics[width=2.3in]{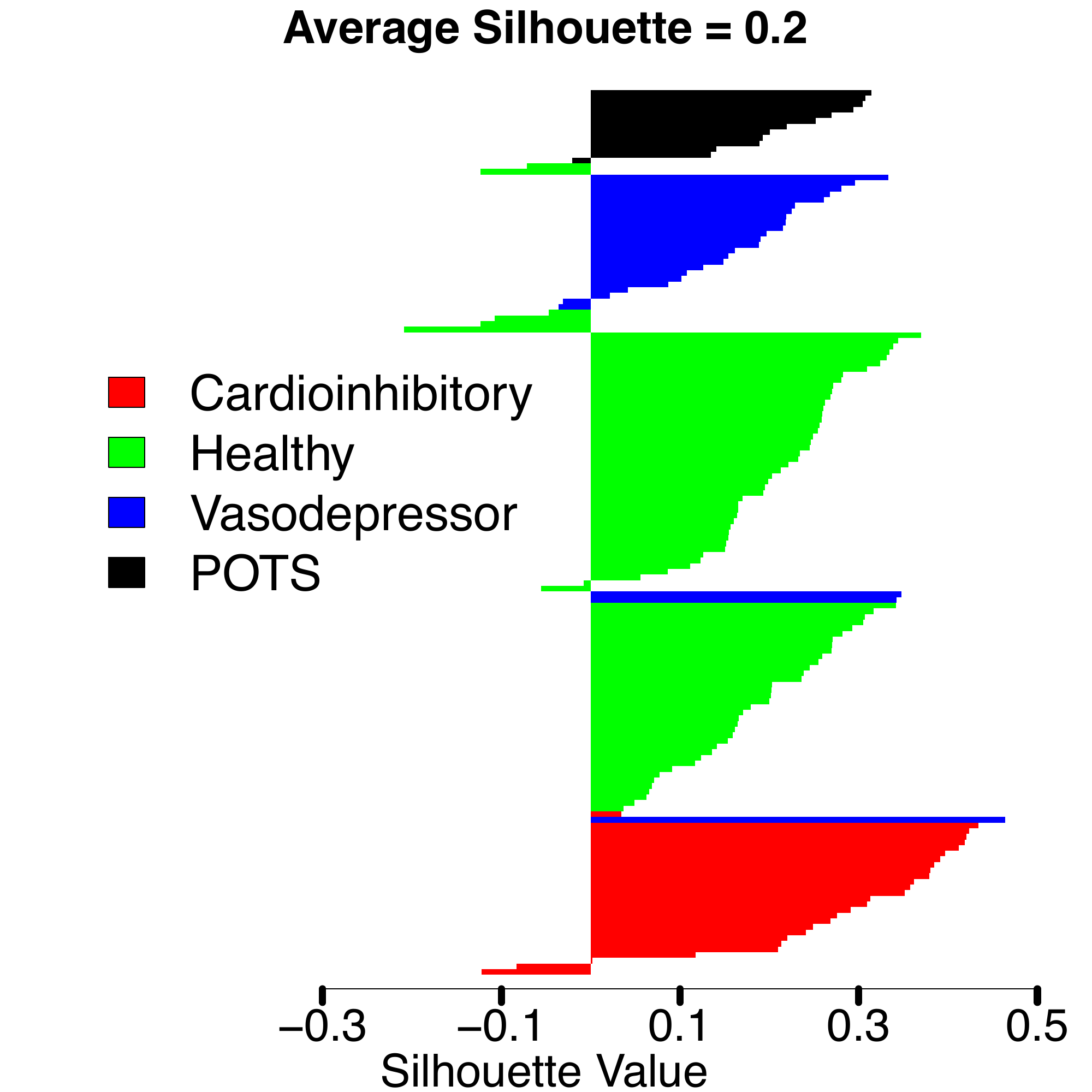}
 \caption{Silhouette representations of the clustering of the population in 4 (left) and 5 (right) clusters.}
\label{fig:clusters}
\end{center}
\end{figure*}

We seek an optimal sampling strategy whereby, within a predefined range, the relative sizes of the intervals defined by $x_1$ and $x_2$ and the number of 
points $z_1$, $z_2$ and $z_3$ in each of them are chosen to maximize classification rate. More precisely, each choice of $\xi = (x_1, x_2, z_1, z_2, z_3)$ defines  a subset of the available data $\mathcal D_\xi$ with 
\[
\mathcal D_{\xi} = \cup_{i=1}^{157} \mathcal D_\xi^i,
\]
where $\mathcal D_\xi^i$ is the subset of the data for the $i$-th subject corresponding to $\xi$.  We construct a cost function through 10-fold cross validation, namely, $\mathcal D_\xi$ is partitioned as follows
\[
\mathcal D_\xi = \cup_{k=1}^{10} \mathcal D_\xi^{\sigma_k} \mbox{ with } \mathcal D_\xi^{\sigma_k} = \cup_{i \in \sigma_k} D_{\xi}^i,
\]
where the $\sigma_k$'s partition  $\{1, \dots, 157\}$. For each $\xi$, we then consider

\vskip .4cm
 \begin{algorithmic}[1]
 \FOR{$k=1$ \TO 10}
 \STATE learn random forest $\mathcal C_\xi^k$ on $\mathcal D_\xi\backslash \mathcal D_\xi^{\sigma_k}$
\STATE compute $r_\xi^k$: classification success rate of $\mathcal C_\xi^k$ on $\mathcal D_\xi^{\sigma_k}$
 \ENDFOR
\STATE $F(\xi) = \frac 1{10} \sum_{k=1}^{10}  r_\xi^k$
 \end{algorithmic}
 \vskip .4cm

The cost function $F$ is  a measure of the successful classification rate as computed through cross validation on the Random Forest model. Note that $F$ inherits the stochasticity of the Random Forest model: two calls to $F$ with the same input parameters may lead to two different outputs. However, the stochastic aspect is mostly negligible here as  classification rates for the same parameterization are observed to change by less than 2\% when the model is run many times.  Experiments show that  10-fold cross validation gives a good approximation of the classification rate attained with leave-one-out cross validation  while allowing for a 20-fold speed-up.

We find the  optimal markers  by solving the  maximization problem 
\begin{eqnarray}
 \underset{\xi}{\operatorname{argmax}} \,F(\xi)  \text{ subject to}\,
\left\{\begin{array}{l}
  0 < x_1 < x_2 < 1, \\
  z_i \mbox{ integer, } i=1,2,3,\\
  z_i \ge 0, i=1,2,3,\\
  12 \le z_1+z_2+z_3 \le 16,
\end{array}\right.
\label{opti}
\end{eqnarray}
where the last constraint was  chosen through trial and error; the retained choice balances  the amount of information and the associated cost. Indeed, to maximize $F$,  we  first fix $z_1,z_2,z_3$ and consider the function mapping from $(x_1,x_2) \mapsto F(x_1,x_2,z_1,z_2,z_3)$ as the objective function. We optimize it using the L-BFGS-B option in the R {\sc{optimx}} function. This is repeated for every possible combination of $z_1,z_2,z_3$ satisfying the constraints. The initial iterate is taken as $(.5,.75)$; numerical convergence is reached in 10 iterations or less in all cases. 
The resulting optimal parameterization is given in Table~\ref{tabopti} and is illustrated in Fig.~\ref{fig:parameterization}.

\begin{table}[h]
\centering
\ra{1.3}
\begin{tabular}{lllll}
\toprule
$x_1$ & $x_2$ & $z_1$ & $z_2$ & $z_3$ \\
\midrule
 0.4999 & 0.9588 & 5 & 7 & 3 \\
\bottomrule
\end{tabular}
\vskip.2cm
\caption{Optimal sampling parameters for (\ref{opti}): 17 nodes are identified.} 
\label{tabopti}
\end{table}

We obtain a total of 17 nodes (with one pre-tilt and one post-tilt nodes) which correspond to 34 markers, 17 BP values and 17 HR values.
Most of the critical information is concentrated immediately before the tilt down time. However, many other parameterizations also attain high success rates. Using the optimal classification rates corresponding to each choice of $z_1,z_2,z_3$ yields 605 parameterizations with mean $93\%$, median $94\%$, min $69\%$ and max $97\%$. We conclude that the classification rate is not sensitive to perturbations in the parameterization.

   \begin{figure*}[!t]
 \begin{center}
\includegraphics[width=.34\textwidth]{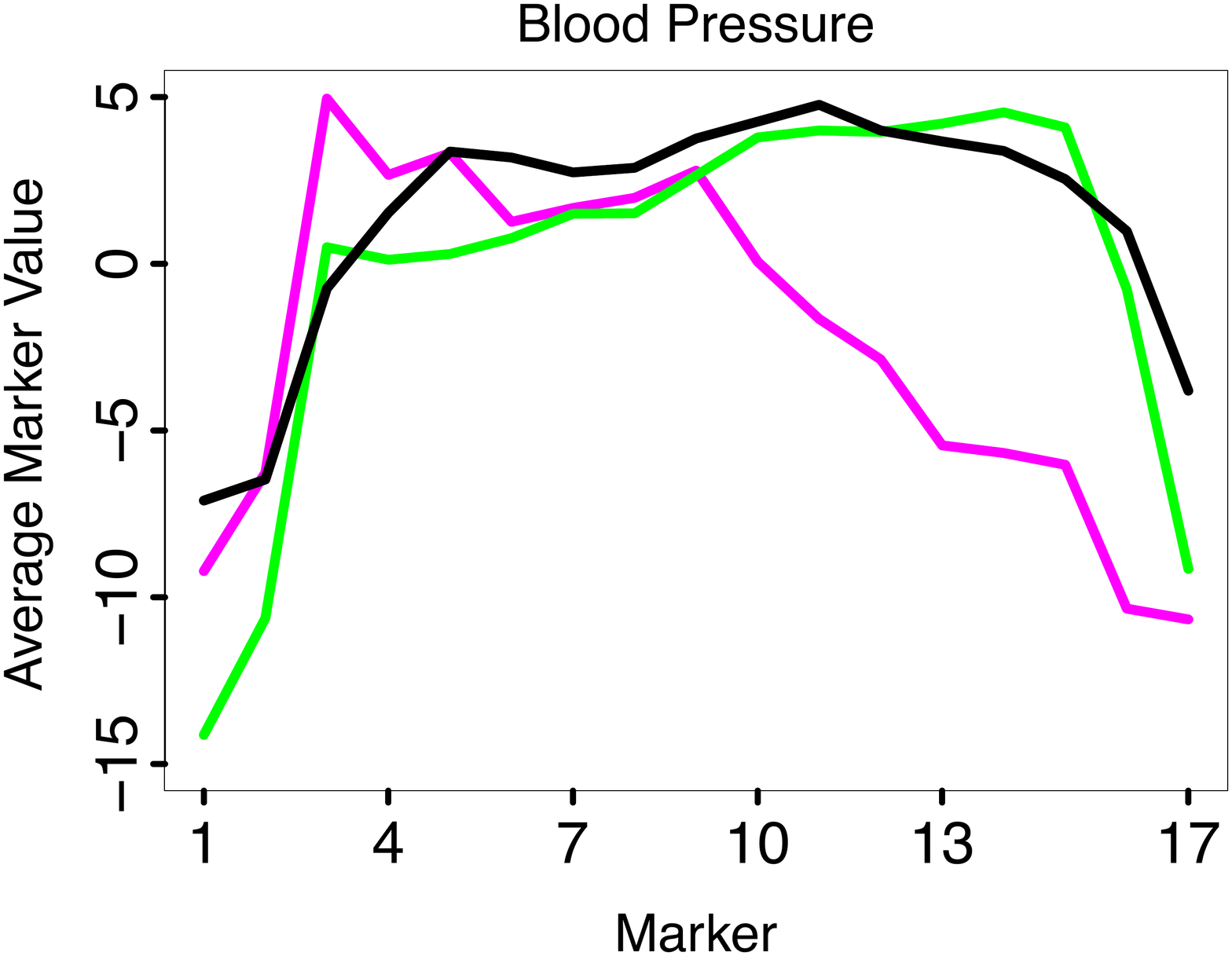}
\includegraphics[width=.34\textwidth]{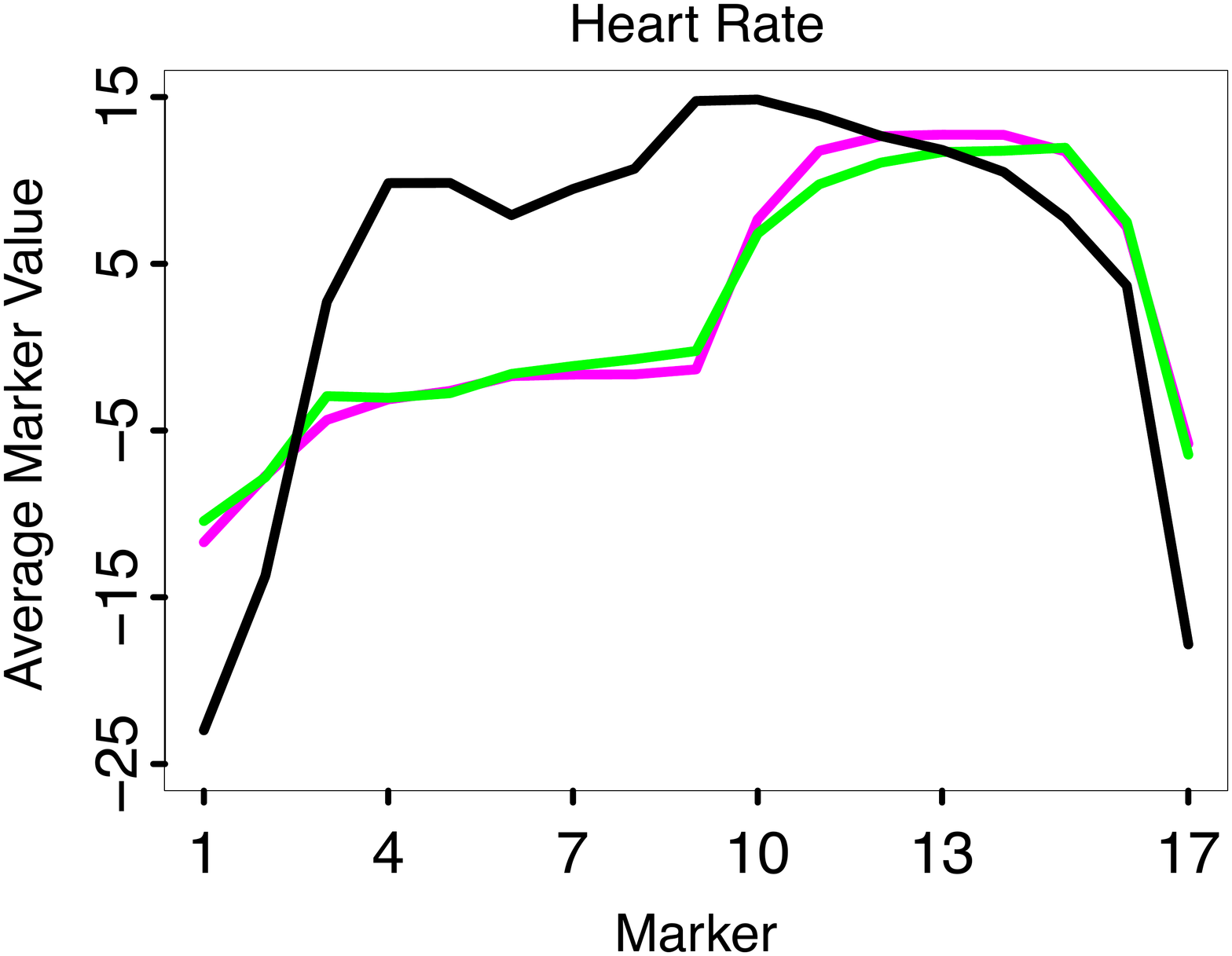}
 \caption{Normalized BP and HR evolution for the two healthy clusters (green and magenta) from Fig.~\ref{fig:clusters}, right,  and the POTS subjects (black); normalized signals are obtained by subtracting the individual global temporal mean from the original signal. Left: BP; the healthy subjects demonstrate two different behaviors; right: HR; the healthy subjects display the same behavior.}
\label{fig:diffbphr}
\end{center}
\end{figure*}

\begin{figure*}[!b]
 \begin{center}
\includegraphics[width=7.5cm]{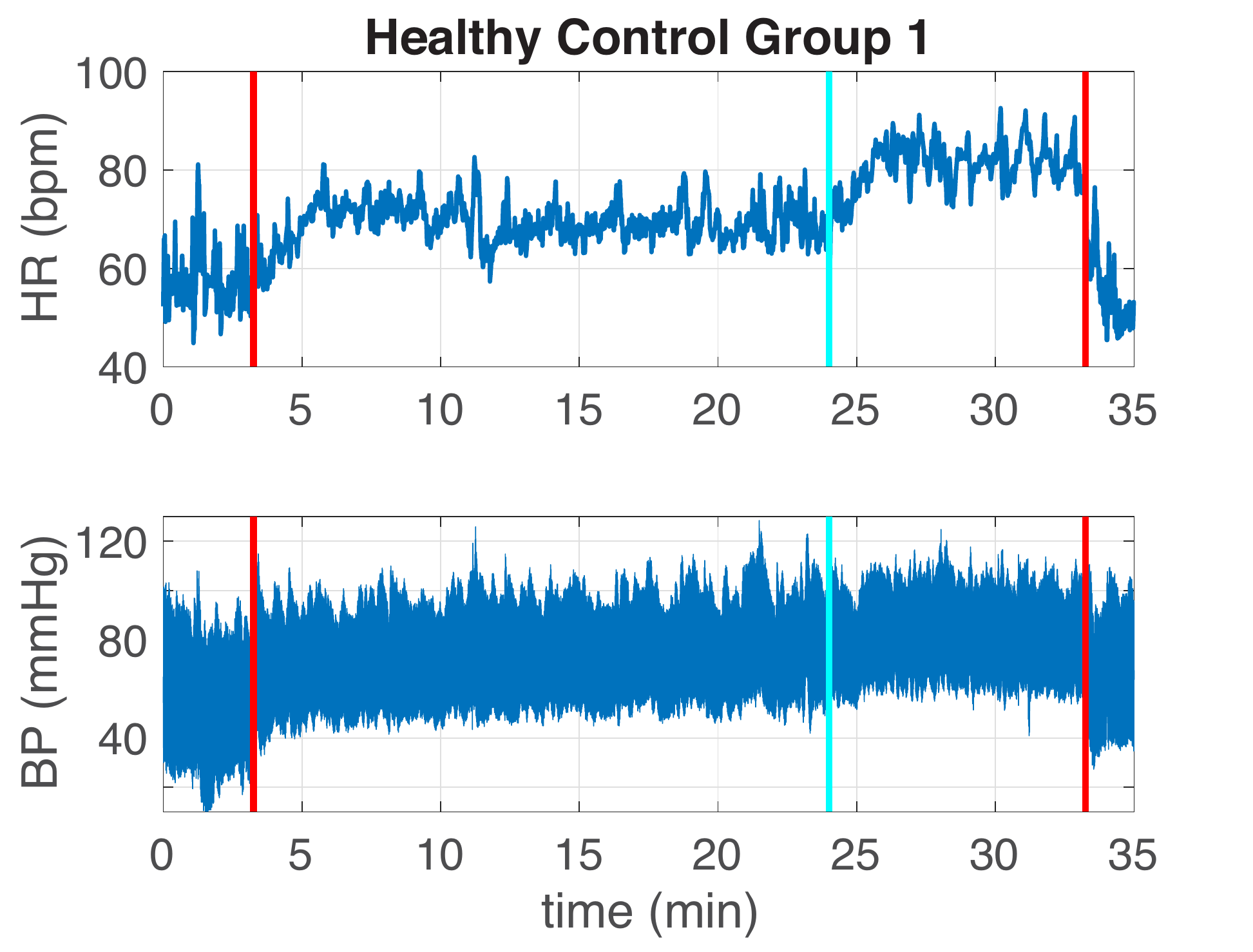}
\includegraphics[width=7.5cm]{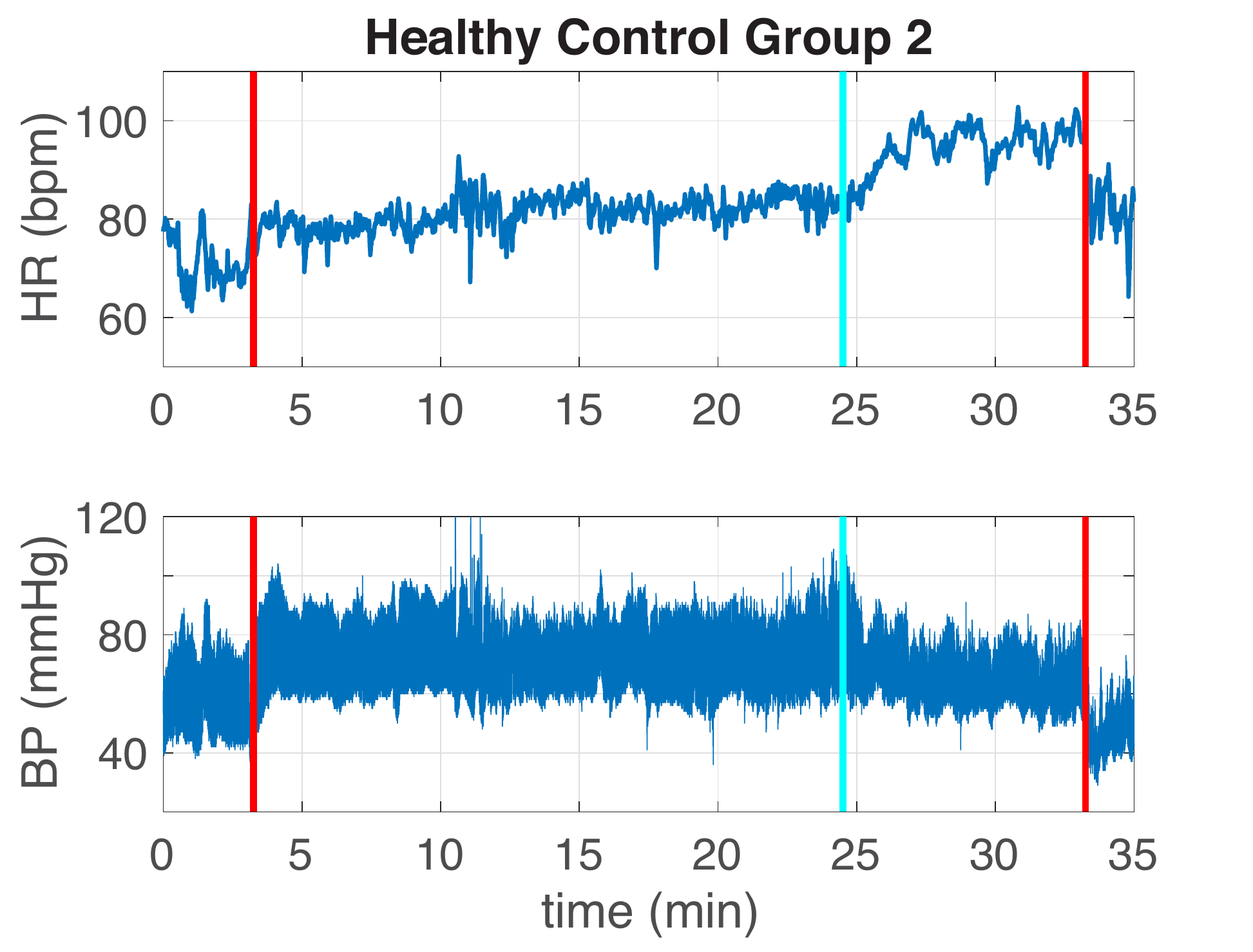}
 \caption{Difference in recorded behavior between a subject from the first group of healthy control (green line in Fig.~\ref{fig:diffbphr}) on the left and a subject from the second group (magenta line in Fig.~\ref{fig:diffbphr}) on the right.}
\label{fig:newfig}
\end{center}
\end{figure*}

\subsection{Clustering}

 We cluster the subjects of the study  through   K-medoids \cite{clarke} \cite{reynolds}, a centroid based clustering algorithm. For that purpose, we use the R implementation  {\sc{pam}} of K-medoids together with the  markers obtained in Section II.C.

The relative importance of these markers can be estimated by permuting out-of-bag data in the Random Forest classification model \cite{liaw}. We denote by $\mathcal I$ the 34-vector of variable importance for these markers. These relative importances are in turn used to  emphasize differences in important variables and facilitate a meaningful clustering process. Specifically, dissimilarities are measured through the matrix $\mathbb D$ with entries
\begin{eqnarray}
\mathbb D_{i,j}=\sqrt{\sum_{k=1}^{34} w_k (m_{i,k}-m_{j,k})^2},\qquad i,j = 1, \dots, 157,  \label{distancec}
\end{eqnarray}
 where $m_{i,k}$ is the value of the $k$-th marker for the $i$-th subject and the weight is given by 
 \begin{eqnarray*}
 w_k=\frac{\mathcal I _k}{\sum_{\ell=1}^{34} \mathcal I_\ell}.
\end{eqnarray*}

\section{Results}

 The Random Forest model determines its classifications according to a majority vote from 500 classification trees. We consider the proportion of votes as a measure of confidence the model has in its classification. Using the optimal sampling strategy from Table~\ref{tabopti}  with leave-one-out cross validation we obtain a classification rate of $96\%$. Fig.~\ref{figclass} shows the patients plotted using the barycentric coordinates of the proportion of votes. The color legend identifies the classification from expert clinicians.

 \begin{figure*}[b]
 \begin{center}
\includegraphics[width=.32\textwidth]{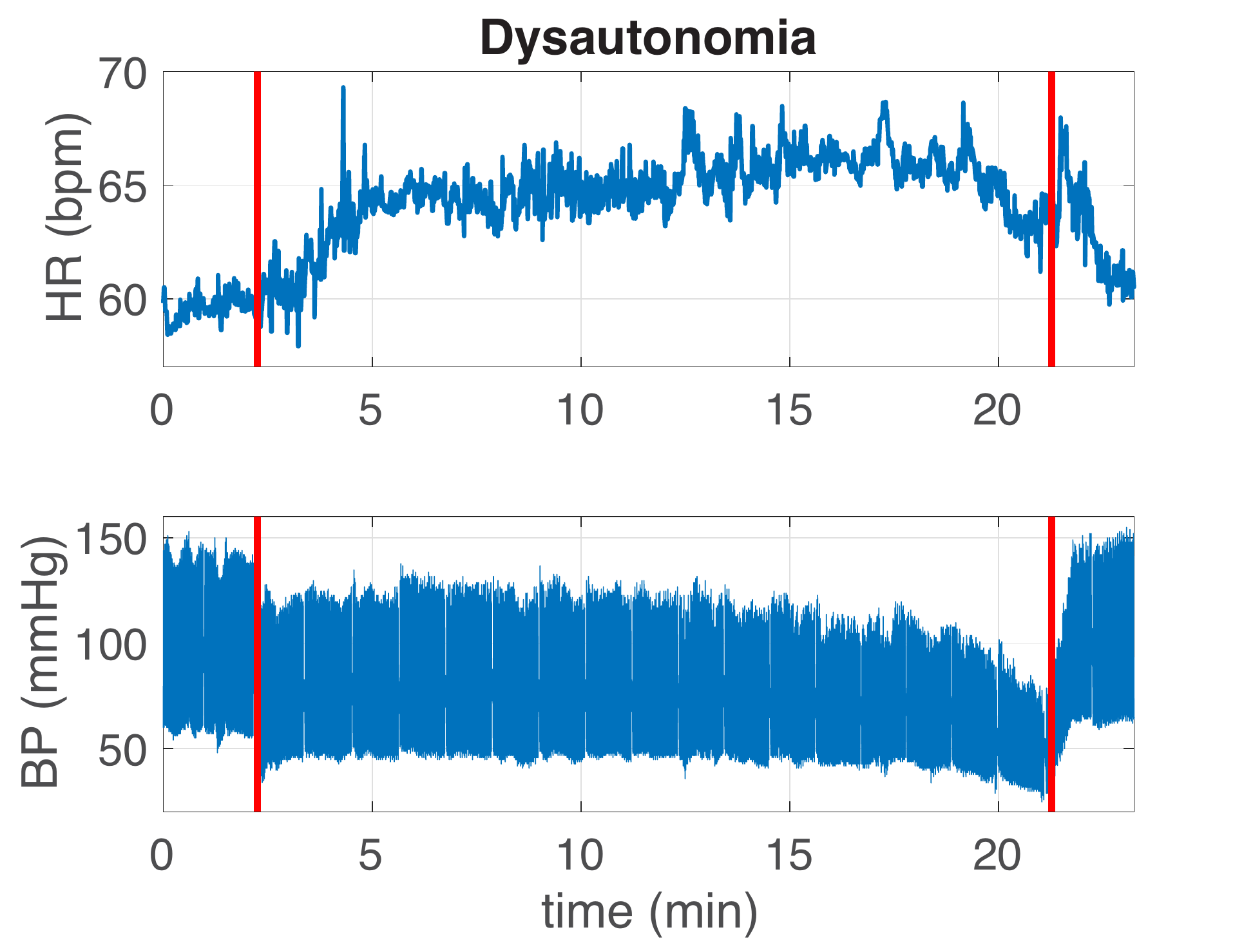} 
\includegraphics[width=.32\textwidth]{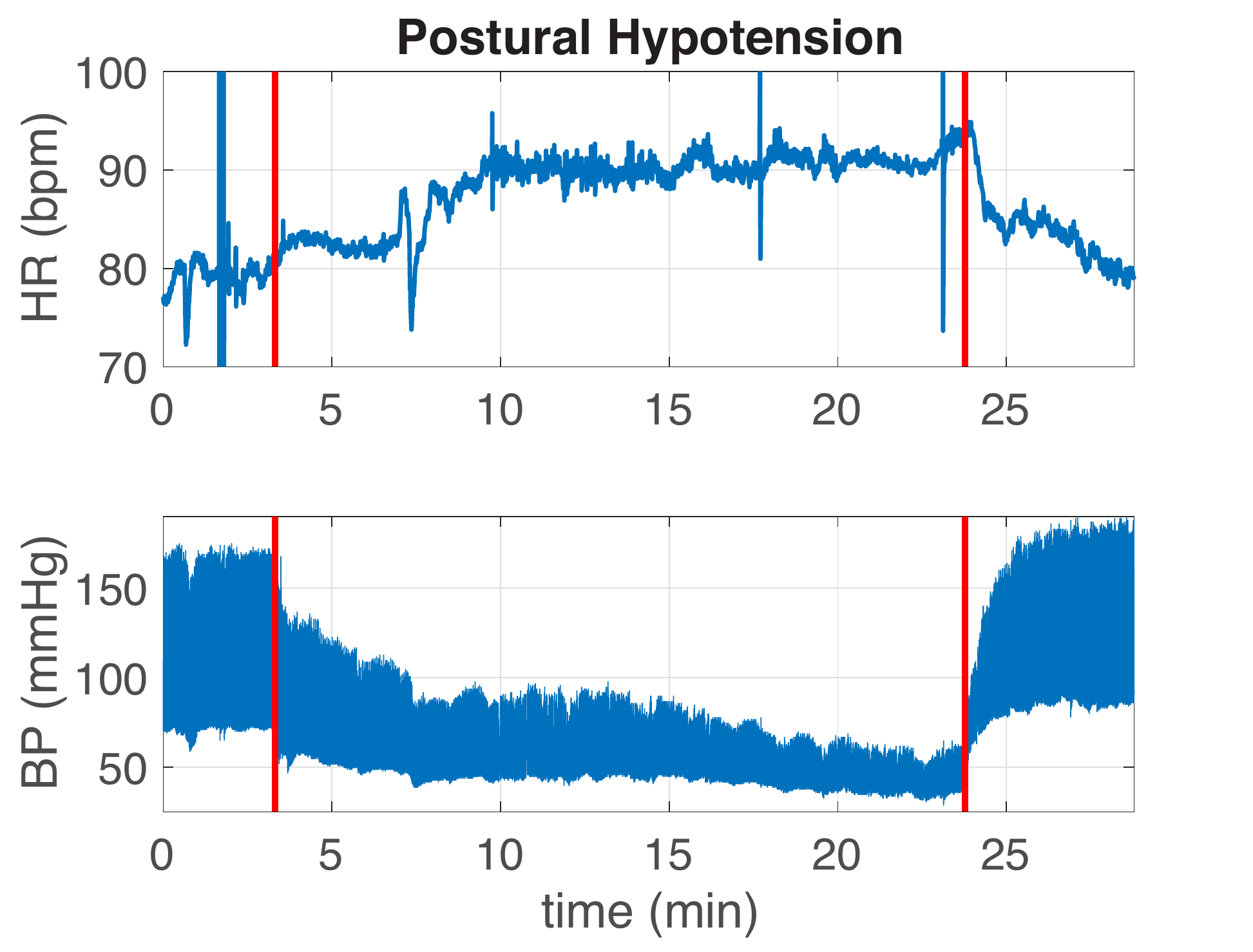} 
\includegraphics[width=.32\textwidth]{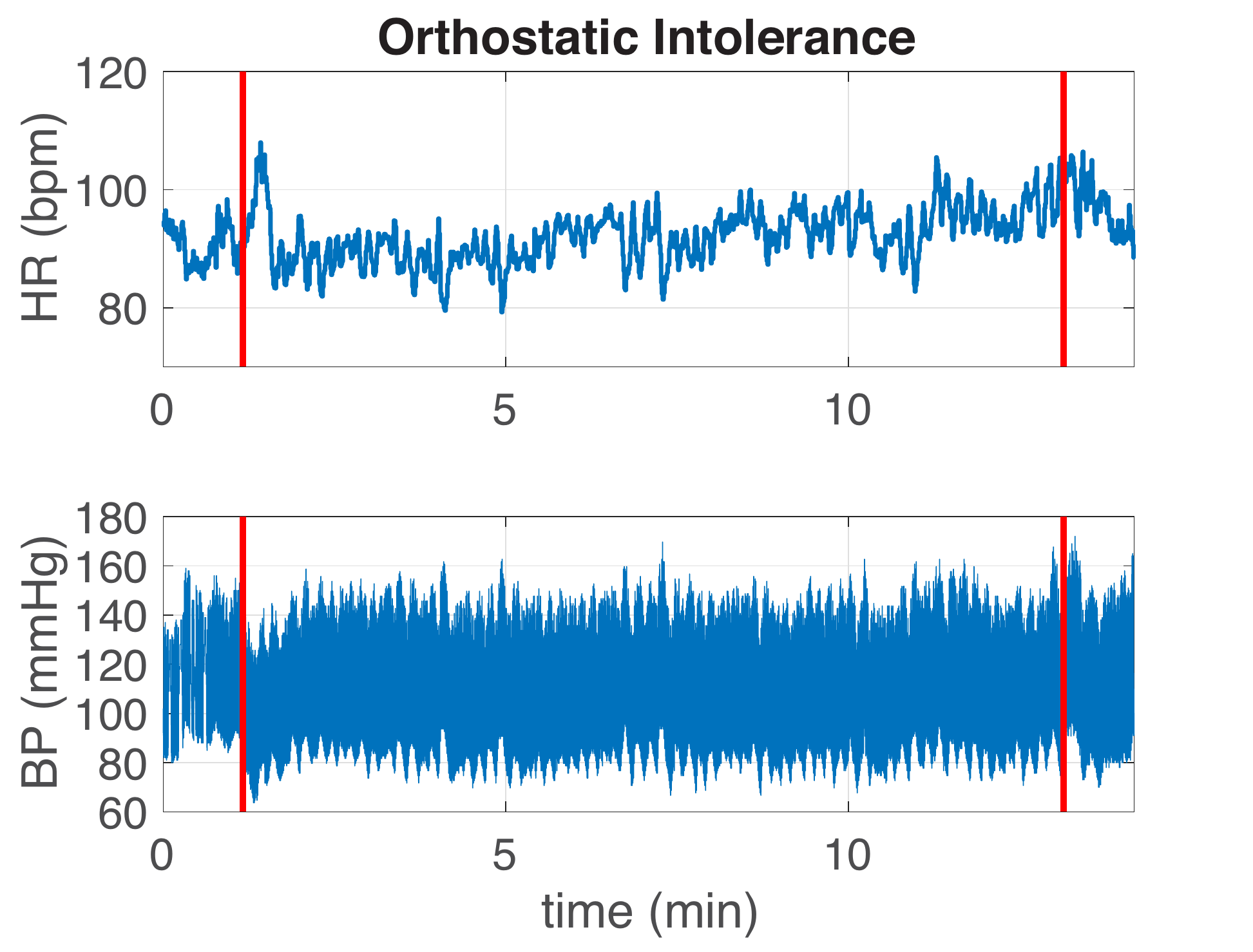} 
\caption{Additional  types of syncope pathologies not included in this project. Left: dysautonomic response. HUT does not lead to a significant increase in HR, likely due to reduced vagal response; sympathetic system regulation may be intact but cannot keep up with the progressive drop in central blood volume due to capillary filtration of fluid from the intra- to the extravascular compartment. Middle: postural hypotension. HUT causes an excessive drop in BP with some change in HR. This could be due to reduced sympathetic vasoconstriction. The small change in heart rate could be caused by intact vagal stimulation. Right: orthostatic intolerance. During HUT, reduced central blood volume causes BP regulation to be challenged by changes in intrathoracic pressure due to respiration. This causes pathological fluctuations in BP with phase-shifted changes in heart rate elicited by the baroreceptor control system.}
\label{fig:morecases}
\end{center}
\end{figure*}

Clustering under the assumption of four distinct clusters leads to  the spreading the healthy subjects into two classes, one with essentially only healthy subjects and the other with a mixture of the rest of the healthy population with the POTS patients. Fig.~\ref{fig:clusters}, left,  displays the Silhouette representation \cite{rousseeuw} corresponding to this clustering. Silhouette values greater than 0 indicate that the patient fits best in its cluster; values less than 0 indicate that it fits better in another cluster. Interestingly, clustering into five groups leads to a surprising results as two different groups of healthy subjects emerge, see Fig.~\ref{fig:clusters}, right,  while the other three groups, i..e, cardioinhibitory syncope, vasodepressor syncope and POTS,  all form their own cluster. We also note that, in agreement with the classification results, see again Fig~\ref{figclass}, the vasodepressor group appears to be the most challenging to characterize.

 Further investigation reveals that there is indeed a distinction between the two identified ``healthy" clusters. This can be seen by computing, across clusters, an average BP at each sample point. The resulting averages are then plotted at each marker, i.e., at a collection in increasing times. In other words, the horizontal axis is a pseudo-time (i.e., a nonlinear time scale).  Fig.~\ref{fig:diffbphr} displays these results for the two healthy cluster of Fig.~\ref{fig:clusters}, right, and the POTS cluster. There is a noticeable difference in BP behavior among the two healthy groups; this separated subjects who experience a drop in BP following nitroglycerine from those who do not not. No such difference is observed for the HR. Figure~\ref{fig:newfig} illustrates data from one subject in each of the two healthy groups.

\section{Discussion and Conclusion}

Based on the above findings, we observe that supervised machine learning--here in the form of Random Forests--can be used to successfully differentiate between healthy subjects and syncope patients; furthermore, our approach can  also identify all three types of syncope considered here (cardioinhibitory, vasodepressor and POTS)  with success rates in the high 90\% among the syncope patients.   Most of the existing related studies concentrate only on differentiating between healthy subjects on the one hand and syncope patients on the other. Various degrees of success are being reported \cite{klemenc} depending on the type of markers/features considered (for instance time domain based versus frequency based), the population size (large versus small), the methods (linear versus non-linear analysis) and the amount of information taken into account. These studies often consider the issue of {\em early} syncope prediction where the goal is to identify subjects susceptible to syncope as early as possible during HUT. While not directly aimed at early prediction, the present work is however relevant to it: the optimal marker locations discussed in Section III clearly (and not surprisingly) emphasize the importance of the information gathered shortly before syncope, i.e., shortly before tilt-down, corresponding to interval 3 above. This is confirmed by \cite{klemenc} where the authors fail to make clinically  useful predictions of the test outcome by concentrating on data from the  first 15mn following tilt-up (and thus mostly ``missing" that critical time). While the results in \cite{virag} are more encouraging, the authors do make use of data in the last minute before syncope in over half of their results.

Our focus is on the multi-class classification and clustering of syncope data.  We are not aware of similar published studies. A possible explanation for the dearth of closely related work might be the difficulty of defining these very classes, a task  the present study starts revisiting. Future work will involve the classification of patients presenting not only the three pathologies discussed above but also other types of syncope such as dysautonomia, postural hypotension and orthostatic intolerance, see Fig.~\ref{fig:morecases}.

Unlike other recent work on syncope data such as \cite{khodor}, we do not retain as features quantities explicitly dependent upon the time-frequency analysis of the two signals BP and HR; instead, we  simply sample the signals at optimized times. Although the inclusion of ``variation dependent features"  did not lead to higher classification rates, we expect that properly chosen quantifiers based on local spectral properties are likely to improve our analysis; this is the topic of ongoing efforts. 

The main purpose of the above classification is the identification of representative markers that can then be used to define a notion of distance (or dissimilarity)  between subjects and, ultimately, for clustering. The distance  between subject $i$ and $j$ is here taken as $\mathbb D_{ij}$ in  (\ref{distancec}). The weighted 2-norm introduced in (\ref{distancec}) is a very natural way of combining the various markers and their relative importance. While clustering largely confirm the validity of the initial clinical classification, it does uncover the existence of two distinct healthy groups. The two healthy groups differentiate patients who are able to maintain BP in response to nitroglycerine versus  those who experience a small BP drop, though not sufficient  to experience pre-syncope or syncope;  all patients in the control group were non-symptomatic (they did not faint). One possible explanation is that the subgroup of healthy controls that experience a BP drop following nitroglycerine administration have sympathetic stimulation operating near or at its maximum (before vasodilation induced by nitroglycerine), and therefore  were not able to maintain a high BP through vasoconstriction in response  to nitroglycerine.

Future work will involve the clustering analysis of patients with symptoms that do not fit the pathologies considered here. Further research is also necessary to investigate possible pathophysiological characterizations of the above two healthy groups. It is  expected that direct mathematical modeling will facilitate the characterization of these and other groups through the testing of  different possible scenarios and root causes.


%

%
%
%

\section*{Acknowledgment}

The authors would like to thank  the Statistical and Applied Mathematical Sciences Institute (SAMSI) where this work was initiated and Peter Novak for helpful discussions.

\ifCLASSOPTIONcaptionsoff
  \newpage
\fi



%

\bibliographystyle{IEEEtran}
\bibliography{prop}

%

%
%
%




\end{document}